\documentclass[%
reprint,
superscriptaddress,
 amsmath,amssymb,
 aps,
]{revtex4-2}

\usepackage{graphicx}
\usepackage{dcolumn}
\usepackage{bm}
\usepackage{xcolor}
\usepackage{siunitx}
\usepackage{gensymb}
\usepackage{amssymb}
\usepackage{hyperref}
\usepackage{cleveref} 
\usepackage{chemformula}
\usepackage{array,booktabs,tabularx, multirow} 

\crefdefaultlabelformat{#2\textbf{#1}#3} 
\crefname{figure}{\textbf{Figure}}{\textbf{Figures}} 

\def\pp{\textit{P.~putida}}
\def\ec{\textit{E.~coli}}
\DeclareSIUnit{\rpm}{rpm}

\begin{document}

\preprint{APS/123-QED}

\title{Swimming patterns of a multi-mode bacterial swimmer \\in fluid shear flow}

\author{Valeriia Muraveva}
\affiliation{%
Institute of Physics and Astronomy, University of Potsdam, D-14476 Potsdam, Germany
}

\author{Agniva Datta}
\affiliation{%
Institute of Physics and Astronomy, University of Potsdam, D-14476 Potsdam, Germany
}
\author{Jeungeun Park}
\affiliation{%
Department of Mathematics, State University of New York at New Paltz, New Paltz, New York 12561, USA
}

\author{Veronika Pfeifer}
\affiliation{%
Institute of Physics and Astronomy, University of Potsdam, D-14476 Potsdam, Germany
}
\author{Yongsam Kim}
\affiliation{%
Department of Mathematics, Chung-Ang University, Seoul 06974, Republic of Korea
}

\author{Wanho Lee}
\affiliation{%
National Institute for Mathematical Sciences, Daejeon 34047, Republic of Korea
}

\author{Sookkyung Lim}
\affiliation{%
Department of Mathematical Sciences, University of Cincinnati, Cincinnati, Ohio 45221, USA
}

\author{Carsten Beta}
\email{beta@uni-potsdam.de}
\affiliation{%
Institute of Physics and Astronomy, University of Potsdam, D-14476 Potsdam, Germany
}
\affiliation{%
Nano Life Science Institute (WPI-NanoLSI), Kanazawa University, Kanazawa, 920-1192, Japan
}

\date{\today}

\begin{abstract}
Bacterial swimming is well characterized in uniform liquids at rest.
The natural habitat of bacterial swimmers, however, is often dominated by moving fluids and interfaces, resulting in shear flows that may strongly alter bacterial navigation strategies.
Here, we study how fluid shear flow affects the swimming motility of the soil bacterium \textit{Pseudomonas putida}, a bacterial swimmer that moves in a versatile pattern composed of three different swimming modes, where the flagella may push, pull, or wrap around the cell body (multi-mode swimmer).
We introduce a computer automated cell tracking and swimming mode detection tool to show that shear induced alignment depends on the swimming mode, while motility and proximity to surfaces counteract the alignment effect.
Moreover, filament wrapping becomes less efficient with increasing shear stress. 
Numerical simulations of realistic swimmer geometries complement our experimental results, providing more detailed mechanistic insights into movement patterns of bacterial swimmers in a shear flow.
\end{abstract}

\maketitle


\section{Introduction}
The locomotion of biological microswimmers in complex environments is central to many medical functions, such as the spreading of infections~\cite{Lauga_2020,  Elgeti2015} or sperm navigation inside the female reproductive tract~\cite{Gaffney2011, ALVAREZ2014}.
For bacteria, one of the largest classes of biological microswimmers, studies in uniform liquid environments have shown that they control their speed and direction of locomotion by changing the mode of operation of their helical flagella.
\textit{Escherichia coli} ({\ec}), the most widely studied example, carries flagella distributed all over its cell body (peritrichous flagellation) and moves in a run-and-tumble pattern~\cite{Berg2004}.
Here, counterclockwise (CCW) rotation of the flagella result in the formation of a coherent bundle that pushes the cell body forward (run). 
Persistent run episodes are interrupted by erratic turn events (tumbles) that are induced by one or several motors switching to clockwise (CW) rotation, driving the coherent bundle apart and resulting in a random reorientation of the cell body~\cite{Turner2000}.

Apart from the well-studied run-and-tumble motility of peritrichously flagellated {\ec}, more complex swimming patterns were reported for bacteria that exhibit other flagellation architectures~\cite{GROGNOT202173}.
While {\ec} always swims as a pusher, other species may exhibit combinations of different run modes (multi-mode swimming), such as, in the run-reverse-flick pattern of monotrichously flagellated species~\cite{Xie2011}.
Here, we focus on the soil bacterium \textit{Pseudomonas putida} ({\pp}) as an example of a multi-mode bacterial swimmer that has been intensely studied both in open uniform liquid~\cite{Theves2013} and in confined and complex environmental geometries~\cite{Theves2015, Raatz2015, Weber2019, Datta2025}. 
{\pp} carries a tuft of several helical flagella located at one cell pole (lophotrichous flagellation)~\cite{Harwood1989}.
It may push itself forward by CCW rotation of its flagellar bundle, while CW operation of the flagellar motor pulls the cell body in the opposite direction.
Additionally, {\pp} can wrap its bundle of flagella around the cell body to move in a screw thread fashion (wrapped mode)~\cite{Hintsche2017e}, a swimming mode that has also been reported for other polarly flagellated species~\cite{Kuhn2017, Kinosita2017, Tian2022}.
Wrapped mode formation is promoted by increasing the viscosity of the surrounding medium~\cite{Kuhn2017,Pfeifer2022a} and plays a key role for chemotaxis of {\pp} in nutrient gradients~\cite{Alirezaeizanjani2020c}. 
Nevertheless, the functional relevance of filament wrapping and the general benefits of switching between different swimming modes are not fully understood and remain a matter of current debate~\cite{Kühn2022}.

\begin{figure*} 
\centering    
\includegraphics[width=1.6\columnwidth]{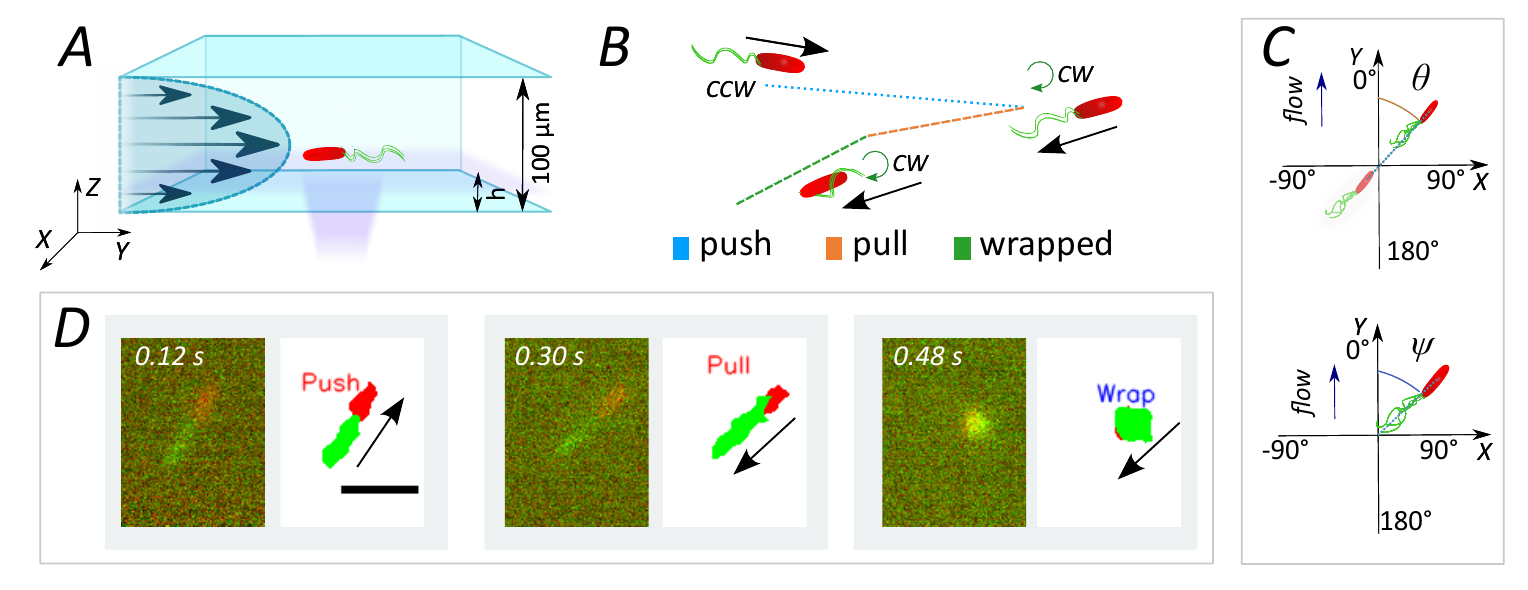}
\caption[Experimental setup and example of analyzed data.]{\textbf{Experimental setup.} (A)~Bacterial suspension in Poiseuille flow in a microfluidic channel. Recordings are taken at a distance \textit{h} above the surface. (B)~Swimming modes of {\pp}. (C)~Angles in the x-y plane: $\theta$ is defined as the orientation of the velocity vector, $\psi$ is the orientation of the cell body, i.e., the vector pointing from the center of mass of the flagellar bundle to the center of the cell body.  (D)~Examples of raw recording data and results of semi-automatic tracking (see the corresponding Video~S1-2 in the Supplementary Information). The arrows show the direction of the velocity vector, scale bar 5~$\mu$m}.
\label{fig:setup}
\end{figure*}

In many natural habitats of bacterial swimmers, fluid flows are commonly observed, such as blood flow in the vascular system, or ground water flows in a granular soil environment~\cite{Rusconi2014, Wheeler2019, Lauga_2020}.
Close to surfaces, the velocity gradients of the resulting shear flows will affect the direction of bacterial locomotion.
According to the pioneering work by Jeffery, a passive elongated object in a linear shear flow will perform periodic revolutions, following so-called \textit{Jeffery's orbits}~\cite{Jeffery1922a}.
The study of shear flow effects on particle dynamics was later extended by Bretherton and others~\cite{Bretherton_1962, Ishimoto2023}.
For active particles in a Poiseuille flow, two types of trajectory patterns, “tumbling” and “swinging”, defined by initial conditions and particle shape, were observed in experiments~\cite{Uppaluri2012} and described theoretically~\cite{Zottl2013}.
Later, also rheotaxis in fluid shear flows, the directed drift of non-motile and motile chiral objects across streamlines, was reported~\cite{Marcos2012b, Zhang2016b, Mathijssen2019c, Jing2020, Zottl2023d} and complemented by mathematical models~\cite{Li2021, Makino2005}.

For elongated swimmers with chiral filaments, the local shear-induced alignment with the flow competes with the chirality-induced side drift~\cite{Zhang2016b, Makino2005}.
Also close to a surface, the combination of shear flow and other hydrodynamic effects leads to the upstream motion of microswimmers~\cite{Kaya2012a, Zhang2016b, Mathijssen2019c}.
All of the above demonstrates that complex environmental flow conditions may strongly affect the spreading of a population of bacterial microswimmers.

In this work, we examine how the swimming behavior of a multi-mode bacterial swimmer is altered under flow conditions.
We focus on {\pp} and report experimental and theoretical results on the swimming pattern of this multi-mode swimmer in bulk and near a solid surface under local shear flow.
To examine the runs in each swimming mode separately, dual color fluorescent microscopy recordings were combined with semi-automatic tracking and swimming mode detection based on a dedicated software tool.
On the theoretical side, these results were complemented with mathematical modeling of a realistic swimmer geometry. The model relied on Kirchhoff rod theory to describe the flagellar filaments and their hooks, rigid body dynamics to capture the motion of the cell body, and a generalization of the regularized Stokeslet method to account for their hydrodynamic interactions~\cite{Kim2022a, Lim2021, Lim2023}.
In this way, we can evaluate the pros and cons of each swimming configuration.
Furthermore, our study will provide the basis for future predictions of the spreading efficiency of biological and novel artificial microswimmers under flow conditions.

\section{Materials and methods}
\vspace{2mm}
\noindent
\textbf{Cell culture and staining.}
%
%
We used the bacterial strain \textit{Pseudomonas putida} KT 2440 $FliC_{\mathrm{S267C}}$ (later in the text referred to as ``wild type"), where the protein FliC was genetically modified for fluorescent staining of the flagella bundle. These bacteria perform run-and-reverse motion in three swimming modes (push, pull, and wrapped, see Figure~\ref{fig:setup})~\cite{Hintsche2017e}. To investigate the influence of the active propulsion, we tested a non-motile strain for comparison. In this strain, the torque-generating stators MotAB and MotCD of the bacterial flagella were knocked out. The {\pp} KT2440 $FliC_{\mathrm{S267C}}$ $\Delta$motAB $\Delta$motCD double knockout mutant was generated by sequential double homologous recombination as described elsewhere~\cite{Pfeifer2022a}.

A day before the experiment, a single colony was harvested from an LB-agar plate and transformed in \SI{25}{\milli\litre} of tryptone broth media (10 g/L tryptone (AppliChem), 5 g/L \ch{NaCl}). The solution was placed in an incubator at \SI{30}{\degreeCelsius} and shaken at \SI{300}{\rpm} for \SI{14}{\hour} (up to a cell density of OD\textsubscript{600} = 0.4). Later, washing with motility buffer (11.2 g/L \ch{K2HPO4}, 4.8 g/L \ch{KH2PO4}, 3.93 g/L \ch{NaCl}, 0.029 g/L EDTA and 5 g/L glucose), the fluorescent staining and further filtration of aggregates were done as described in~\cite{Pfeifer2024}. In short, a washed concentrated bacterial suspension was incubated with Alexa Fluor 488 C5-maleimide (Invitrogen, Thermo Fisher Scientific) over \SI{30}{\min} at room temperature for staining of the flagella. After that, the suspension was washed and re-incubated with the red membrane dye FM4-64 (Invitrogen, Thermo Fisher Scientific) to visualize the cell body. During the last step of staining, the bacterial suspension was gently filtered to remove aggregates of cells that appeared during intermediate centrifugation and washing steps.
Finally, the cell suspension was resuspended to OD\textsubscript{600} $\sim$ 0.2 (low-density suspension of cells, to avoid cells-cells interaction) in motility buffer, which forced the bacteria to stop growing but kept them motile. 

\vspace{2mm}
\noindent
\textbf{Microfluidic setup and image acquisition.}
%
%
The suspension of fluorescently stained cells was injected with a syringe into a rectangular polymer channel of type $\mu$-Slide 0.1 VI, ibiTreat (Ibidi), with height H =\SI{100}{\micro\metre} (the setup is shown in Figure~\ref{fig:setup}A). For experiments under flow conditions, the bacterial suspension was pushed through the channel with a constant flow rate by a syringe pump (Harvard Apparatus, 11 Elite). To achieve equilibrium in the system, the pump was running \SI{20}{\min} before the first measurements. For the bulk experiments, the flow rates were 250 nl/min or 500 nl/min, resulting in local shear rates of $\gamma =1.5$ s$^{-1}$ or $\gamma=3.0$ s$^{-1}$, respectively, at a distance of $h = \SI{20}{\micro\metre}$ above the surface. For $h = \SI{5}{\micro\metre}$ above the surface, the flow rate was decreased to 125 nl/min to maintain a shear rate of $\gamma=3.0$ s$^{-1}$. The experiments with non-motile mutant cells were carried out at $h = \SI{20}{\micro\metre}$ and under a local shear rate of $\gamma=3.0$ s$^{-1}$.

For fluorescence imaging, an inverted microscope (Olympus, IX71) was equipped with a monochromatic blue LED source with $\lambda$ = \SI{470}{\nm} (Prizmatix, UHP-T-LED-470). For splitting the red signal of the cell body and the green signal of the flagella, the splitting optics (Hamamatsu, W-View Gemini) were used. 
The recordings were taken with a CMOS high-speed camera (Hamamatsu, ORCA-Flash 4.0 LT, C15440-20UP) in a fixed focal plane at distances of \textit{h}~=~\SI{5}{\micro\meter}, or \SI{20}{\micro\meter} above the bottom of the microfluidic channel (see Figure~\ref{fig:setup}A). Images were acquired with a time resolution of 100 frames per second, using a 60x UPLFLN-PH objective (Olympus) with a small depth of field (DP = \SI{0.7}{\micro\metre}) to obtain information from a thin layer of the sample.

\vspace{2mm}
\noindent
\textbf{Image processing and cell tracking.}
%
%
To track and distinguish cells in different swimming modes, bacteria were stained with fluorescent dyes (red for the body and green for the flagella, see Paragraph on Cell Culture and Staining above and Figure~\ref{fig:setup}B). Preliminary processing of the fluorescence recordings (channel splitting and frame cropping) was done with the freely available ImageJ software as described earlier~\cite{Hintsche2017e}. For further tracking and analysis, a custom-made Python code was used~\cite{Datta_code2025}, see also Supplementary Information. 

\begin{figure} [h]
    \centering
    \includegraphics[width=0.8\linewidth]{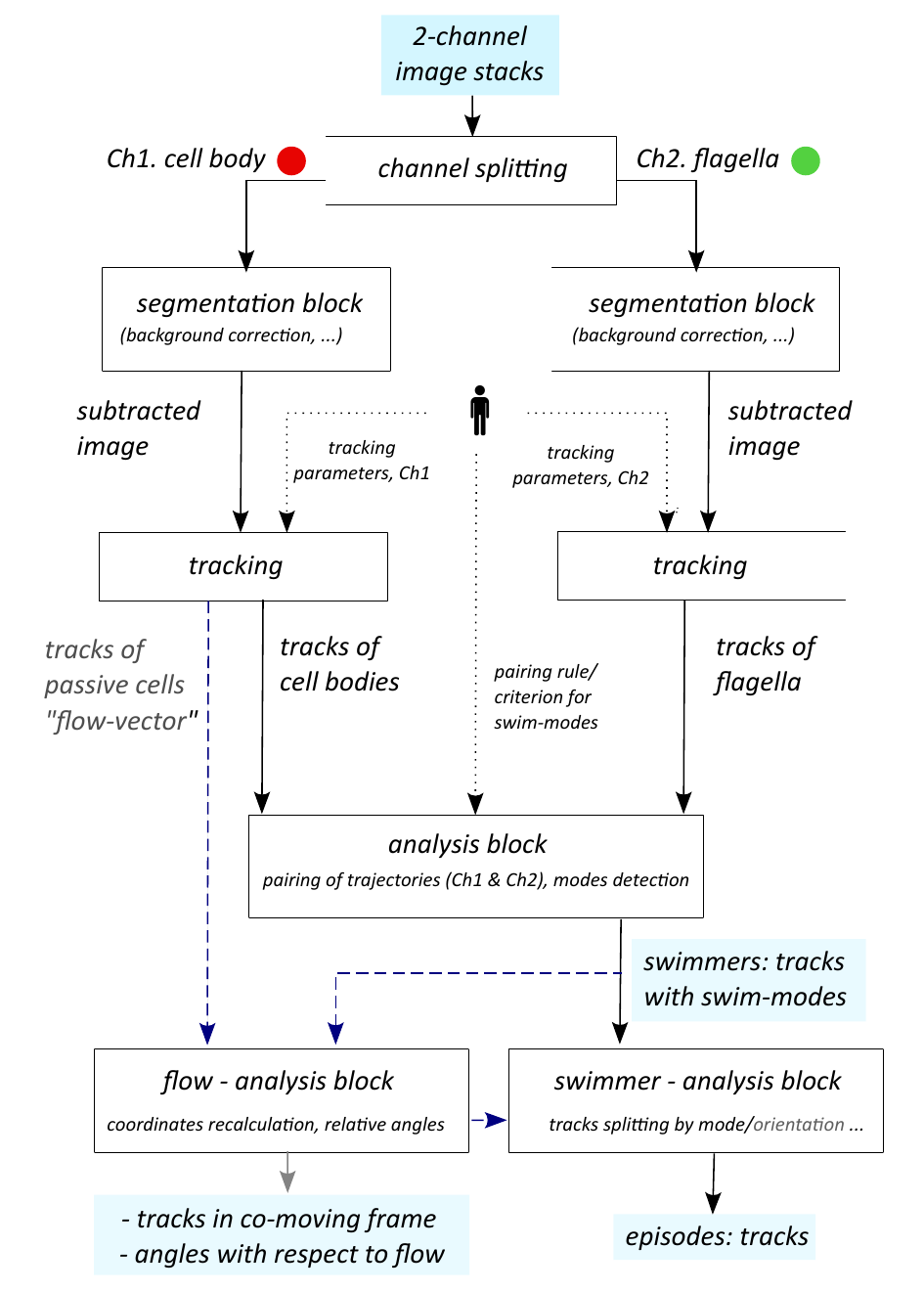}
    \caption{ \textbf{Activity diagram of the imaging analysis workflow.} In the analysis block, the tracks of swimmers are considered as detected from the positions of the swimmers' cell bodies unless otherwise stated. With the blue dashed arrows, the steps required for the flow subtraction are shown. We use the human icon and dotted arrows to mark the parameters that a user must specify manually.}
    \label{fig:code_diagram}
\end{figure}

The two channels corresponding to the fluorescence recordings of the red and green signals were segmented and tracked separately, see the workflow scheme in Figure~\ref{fig:code_diagram}.   
For every position in a trajectory in the red channel (associated with the center of mass of the cell body), the position of the corresponding trajectory in the green channel (associated with the center of mass of the flagellar bundle) of the same bacterium was determined based on a search radius, which was taken to be comparable to the size of the bacterium.
The two points were then connected by a vector.

To subtract the flow velocity, separate recordings with only non-motile cells were used for each measurement condition. 
The speed of well-resolved inactive cells was averaged and subtracted from the velocities of the cell bodies of the active swimmers obtained from the analysis of trajectories in the red channel. To distinguish between the swimming modes, the orientation of the vector pointing from the center of mass of the flagella to the center of the cell body with respect to the direction of cell displacement was analyzed, see Figure~\ref{fig:setup}C. 

If the vector connecting flagella and cell body was pointing in the same direction as the displacement vector (after subtracting the fluid flow speed), the cell was assigned to be a pusher. On the other hand, a puller was identified when these vectors were oriented in opposite directions. Finally, if the distance between the centers of mass of the cell body and the flagellar bundle fell below a threshold of 1.1~$\mu$m, the swimmer was assigned to move in wrapped mode (see also Table S1 in Supplementary Information). Prior to the swimming mode detection, trajectories were split into individual run episodes, for which the swimming modes were then identified separately, see also Figure~\ref{fig:setup}B.

Note that this analysis relies on two types of orientation angles (see Figure~\ref{fig:setup}C). The angle $\psi$ characterizes the orientation of the cell body and is defined as the angle between the vector, which connects the centers of mass of the flagellar bundle and the cell body, and the direction of the fluid flow velocity, determined from the drift direction of non-motile cells. The angle $\psi$ can be determined for both pushers and pullers, but not for wrapped swimmers, due to the overlap of a cell body and a flagellar bundle. Also, for the passive non-motile mutant cells, $\psi$ can be determined.

On the other hand, the angle $\theta$ represents the orientation of the swimming velocity vector with respect to the direction of fluid flow. The velocity vector is determined by connecting the cell body positions in adjacent time frames, see also Figure~\ref{fig:setup}C.

\vspace{2mm}
\noindent
\textbf{Mathematical framework.}
%
%
To simulate the hydrodynamics of a swimming bacterial cell in a viscous fluid, we employ the regularized Stokeslet formulation coupled with Kirchhoff rod theory, implemented within the immersed boundary framework, which is well suited for fluid-structure interaction problems~\cite{Lim2013,Lim2021}. The cell body is modeled as a neutrally buoyant, rigid body and the flagellum is treated as an elastic slender filament governed by Kirchhoff rod theory. The flagellum is attached to the pole of the cell body to complete the full cell model and this cell is immersed in an incompressible Newtonian fluid. The bacterial flagellar motor generates torque that drives the rotation of the flagellum. Simultaneously, an equal and opposite torque is applied to the cell body, resulting in its counterrotation. The swimming motion is calculated by enforcing force-free and torque-free conditions on the surrounding fluid. This framework enables the accurate capture of both direct mechanical interactions and fluid-structure coupling between the cell body and its motor-driven, flexible flagella. See Supplementary Material for a detailed mathematical description.

\section{Results}
\begin{figure}[t] 
\centering    
\includegraphics[width=1.0\columnwidth]{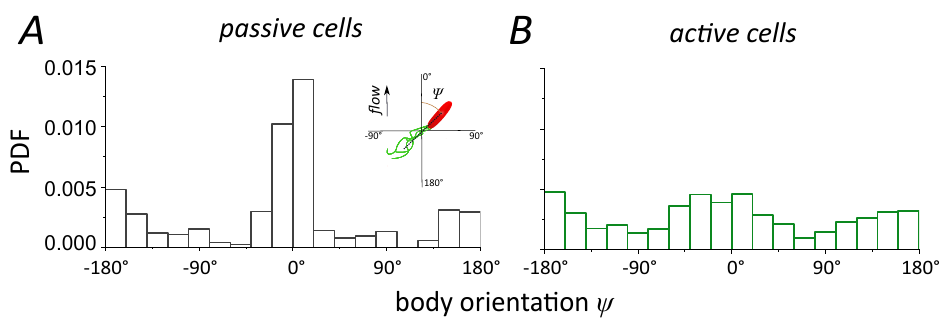}
\caption[Angles mutant]{\textbf{Flow alignment of cell bodies of motile and non-motile bacteria.}
Distribution of body orientations ($\psi$ angles) of (A)~non-motile mutant cells and (B)~motile wild type cells with respect to flow. Only unwrapped configurations (push and pull for swimming cells) were considered; $\gamma=3.0$~s$^{-1}$, $h=\SI{20}{\micro\metre}$.}
\label{SI_fig:mutant angles}
\end{figure}

\noindent
\textbf{Motility counteracts alignment of bacteria in shear flow.}
It is well known that the orientation of elongated objects in a moving fluid is affected by shear flow conditions~\cite{Jeffery1922a, Elgeti2015, Ishimoto2023, Zottl2019, Rubio2021}.
Here, we analyse the impact of shear flow on the alignment of motile {\pp} cells.
We first elucidated the overall impact of motility on the alignment of rod-shaped, lophotrichous flagellated bacterial cells in a shear flow. We compared the cell body orientations ($\psi$ angles) of motile {\pp} wild type cells with non-motile {\pp} mutant cells.
A $\Delta motAB$ $\Delta motCD$ double mutant, deficient in both stators of the flagellar motor, was used as the non-motile reference case, as these cells still carry flagellar filaments, but are unable to rotate their flagellar motors\cite{Pfeifer2022a}.
For both wild type and non-motile cells, alignment with the flow direction was observed, indicated by peaks at $0^{\circ}$ and $180^{\circ}$ in the orientation angle $\psi$ histograms displayed in Figure~\ref{SI_fig:mutant angles}.
However, as can be seen from the heights of the peaks in the histogram, the alignment effect was weaker for the motile wild type as compared to the non-motile mutant cells, where the alignment peaks were more pronounced. 

\begin{figure}[b] 
\centering    
\includegraphics[width=1.0\columnwidth]{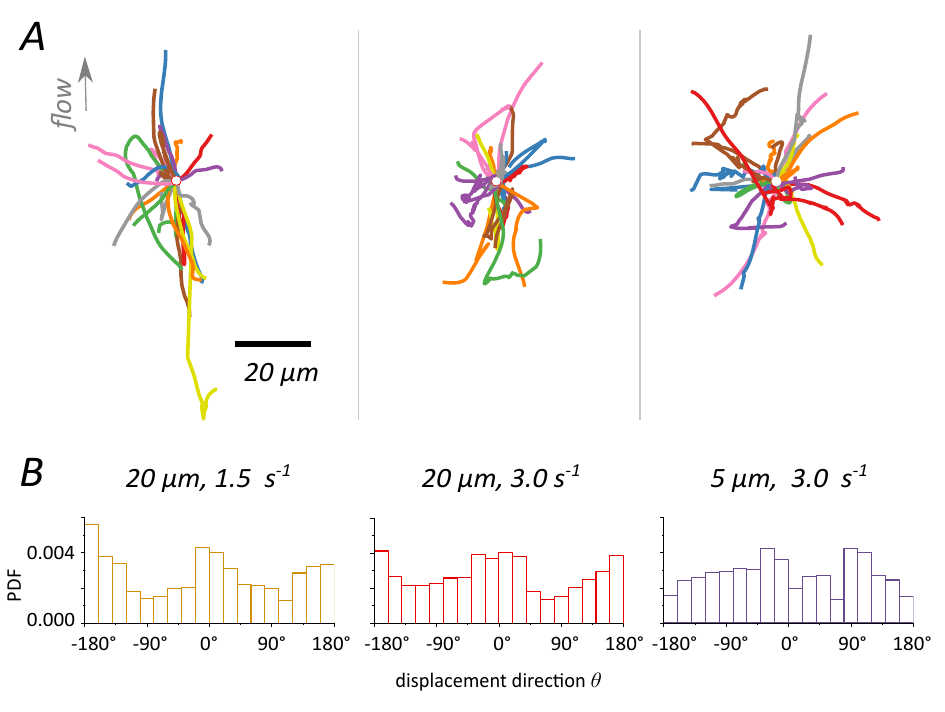}
\caption[Surface and flow, tracks]{\textbf{Alignment of bacterial locomotion.} (A)~Typical swimmers’ trajectories in the co-moving frame, 25 longest tracks are shown. Flow speed is subtracted, flow direction is parallel with the Y-axis, and the beginning of the tracks is marked by a white circle.
(B)~Distribution of $\theta$ angles (swimming direction) with respect to the flow in bulk ($h=$~\SI{20}{\micro\metre}) and near the surface ($h=$~\SI{5}{\micro\metre}). The time lag is $\Delta T=0.05$~s.}
\label{fig:tracks_flow}
\end{figure}
\begin{figure*} 
\centering    
\includegraphics[width=1.65\columnwidth]{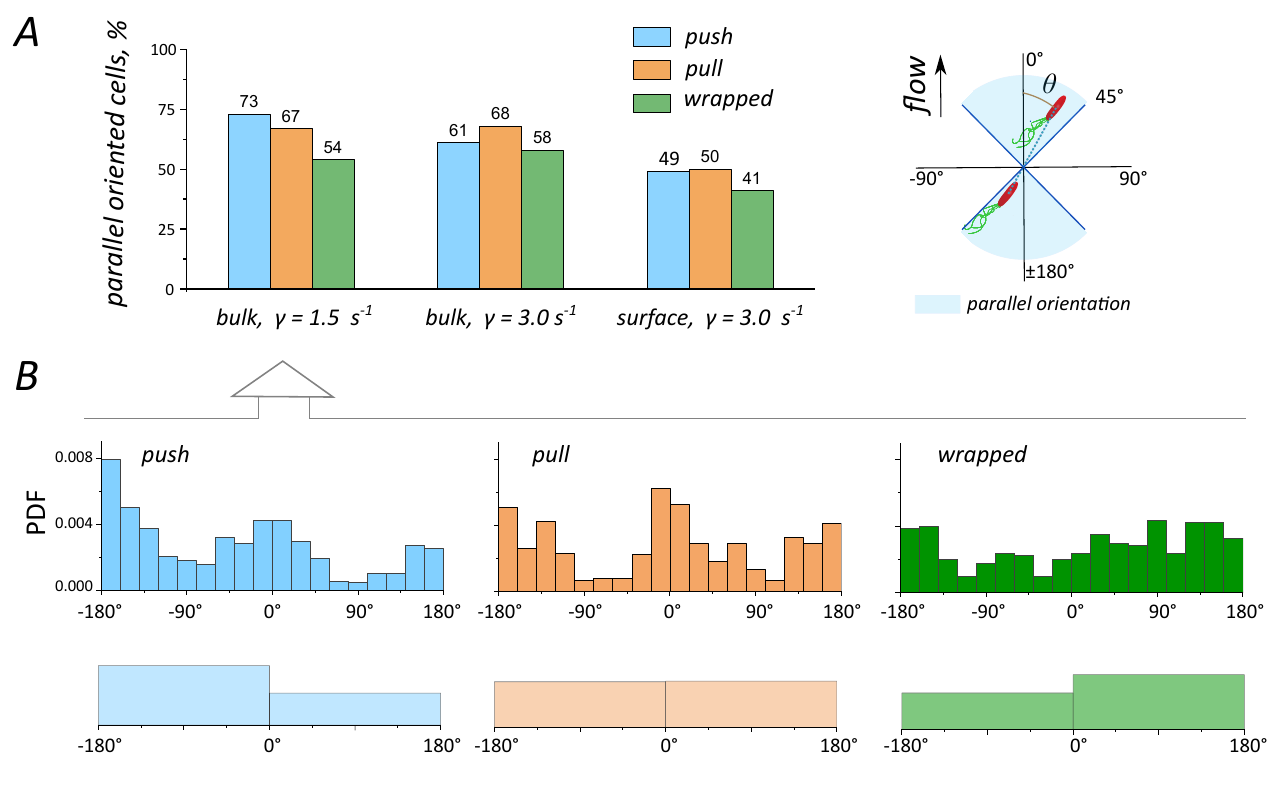}
\caption[Angles, bulk]{\textbf{Flow alignment of swimming modes.} (A)~Percentage of displacements parallel to the direction of fluid flow, i.e. within a deviation from streamlines of less than $\pm$45$^{\circ}$, see insert on the right; ``bulk" refers to $h=$~\SI{20}{\micro\metre}, ``surface" refers to $h=$~\SI{5}{\micro\metre} above bottom surface of the channel.
(B)~Distributions of $\theta$ angles is separated according to swimming modes ($h=$~\SI{20}{\micro\metre}, $\gamma= 1.5 $~s$^{-1}$). The time lag is $\Delta T=0.05$~s.}
\label{fig:Angles_bulk}
\end{figure*}

\vspace{2mm}
\noindent
\textbf{Shear-induced alignment of swimming trajectories is reduced close to surfaces.}
Even though flagellar activity compromises alignment in a shear flow, a clear preference to orient with the flow direction was observed. This was also reflected at the level of cell trajectories. In Figure~\ref{fig:tracks_flow}A, the swimming trajectories of {\pp} wild type cells are shown for two different shear rates in the bulk fluid (at a distance of \SI{20}{\micro\metre} away from the bottom of the microchannel) and in a shear flow close to a surface (at a distance of \SI{5}{\micro\metre} from the bottom of the microchannel).
The trajectories are displayed in a co-moving reference frame, i.e., the speed of fluid flow in the focal plane was subtracted, and their starting positions were centered at the origin.
For the two bulk fluid cases (left and middle), the pattern of trajectories is extended in flow direction, indicating an alignment of the direction of motion with the fluid flow.

This is also reflected in the histogram of displacement angles $\theta$ shown in Figure~\ref{fig:tracks_flow}B.
For both shear rates, similar peaks at $0^{\circ}$ and $180^{\circ}$ indicate a preferred swimming direction with or against the fluid flow.
In contrast, alignment vanished in the vicinity of the surface, see Figure~\ref{fig:tracks_flow}, right-hand side.
Here, the orientations of trajectories are more isotropic and no clear peaks can be distinguished in the histogram of the displacement angle.

\vspace{2mm}
\noindent
\textbf{Pushers and pullers align more strongly than swimmers in wrapped mode.}
Thus, in a fluid shear flow away from solid boundaries, we observed an overall alignment of swimming trajectories with the flow direction that breaks down close to a solid boundary.
In order to find out to what extent alignment depends on the swimming mode, we performed dual color fluorescence imaging experiments allowing us to identify for each run, whether the respective cells move in push, pull, or wrapped mode (see Figure~\ref{fig:setup}D relying on our semi-automated tracking and assignment software, see Materials and Methods).
Based on the assignments of runs to the different swimming modes, we quantified the orientations of cell displacements relative to the flow direction for each swimming mode separately.
We classified a displacement as being aligned with the flow, if its orientation deviated from the flow by less than $45^{\circ}$ (with or against the flow direction, see schematic in Figure~\ref{fig:Angles_bulk}A).
The bar plots in Figure~\ref{fig:Angles_bulk}A display the percentages of displacements aligned in flow direction for each swimming mode separately.
The three different flow configurations correspond to the cases shown in Figure~\ref{fig:tracks_flow}.

The bar plots confirm that in a shear flow far from solid surfaces, swimming directions preferentially align with the fluid flow, whereas alignment vanishes at the surface.
With respect to the different swimming modes, these results moreover indicate that push and pull modes have an increased tendency to align as compared to cells swimming in wrapped mode.

We also considered the temporal reorientation dynamics of individual swimmers (see Figure~S4 in Supplementary Material).
For long episodes in push or pull mode, the $\theta$ angle mostly fluctuates around 180$^{\circ}$ (against flow direction). Cells in wrapped mode, in contrast, reorient faster and undergo more pronounced changes in their swimming direction.
Note, however, that only small numbers of sufficiently long trajectories were available in our data, so that we cannot provide a rigorous statistical analysis of the reorientation dynamics.

\begin{figure*}
\centering    
\includegraphics[width=1.85\columnwidth]{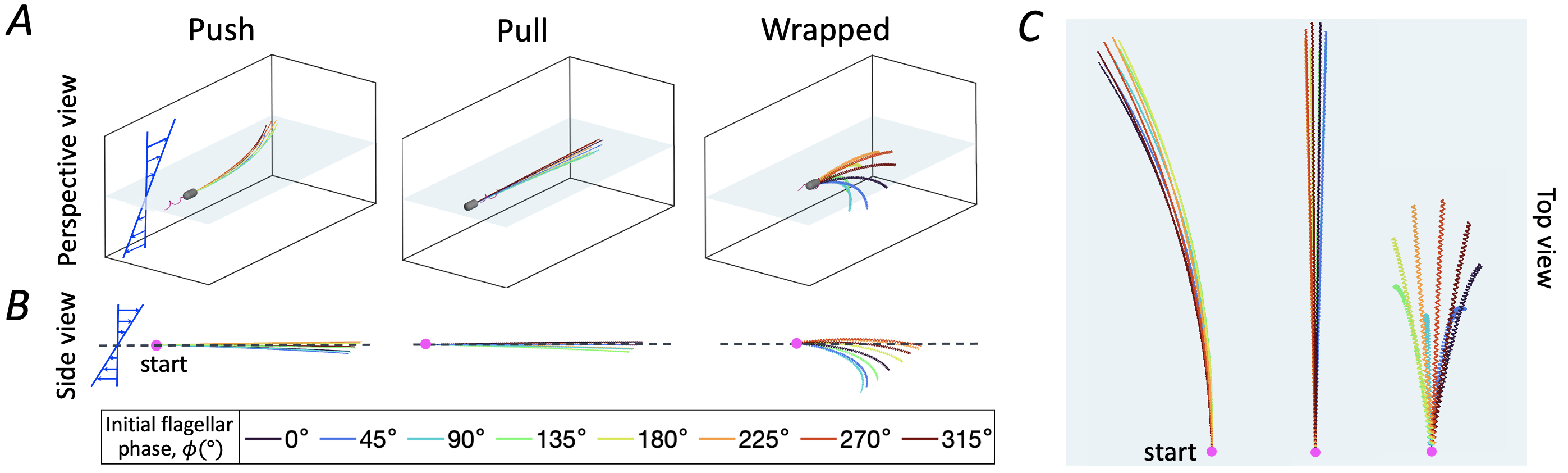}
\caption[Surface and flow, tracks]{\textbf{Numerical simulations: trajectories of bacteria in a shear flow.}
Trajectories of swimmers in push, pull, and wrapped modes under shear flow conditions. (A)~Perspective, (B)~side, and (C)~top views are shown for each swimming mode. In each view, eight trajectories, corresponding to eight different initial phases, are displayed for each swimming mode. In the light blue plane, the flow velocity is zero. Each simulation is run for 0.8~s with a shear rate of $\gamma =3.0$ s$^{-1}$. The torque values for each mode are as follows: $\tau=2.0$ nN$\cdot$nm for push, $\tau=-2.0$ nN$\cdot$nm for pull, and $\tau=-4.0$ nN$\cdot$nm for the wrapped mode.
Above the plane of focus, the fluid moves in line with the initial swimming direction (swimming with the flow), while below it moves against the initial swimming direction (swimming against the flow).
See also the corresponding Video~S3.}
\label{fig:tracks_model}
\end{figure*}

In Figure~\ref{fig:Angles_bulk}B, we consider the full histograms of displacement angles $\theta$ of all three swimming modes for a shear flow of $\gamma=1.5$~s$^{-1}$ away from a solid surface.
In agreement with the bar plots in panel~A, the histograms of push and pull modes exhibit pronounced peaks at $0^{\circ}$ and $180^{\circ}$, indicating alignment in flow direction, while no clear pattern is observed in the histogram of the wrapped mode.
Interestingly, the histogram for {\pp} swimmers in push mode additionally shows a preferred directional drift across streamlines in $-x$ direction (see also Figure~\ref{fig:setup}A), which means towards the right-hand side with respect to the negative flow ($-y$) direction. This can be seen as an indication of bacterial rheotaxis in bulk resulting from an interplay between the helical flagellar bundle and the velocity gradient of the fluid shear flow~\cite{Marcos2012b,Mathijssen2019c}.
Even though a similar chiral flagellar tuft is present in both pushers and pullers, contrary to our expectations, no drift was observed for runs in pulling mode.
However, the compact wrapped configuration of the swimmer, which does not exhibit pronounced alignment with the streamlines, shows a directional bias to the opposite side, which, together with the behavior of pushers and pullers, will be considered in the following numerical simulations.

\vspace{2mm}
\noindent
\textbf{Numerical simulations confirm different rheotactic responsiveness of pushers and pullers.}

To further substantiate whether {\pp} cells swimming in push and pull modes show different sensitivity in their rheotactic response to shear flow, we performed numerical simulations. 
Our computational cell model is based on the real geometry of {\pp}, with the exception that the cell has a single superflagellum at one pole. This simplification assumes that all flagella, when driven by a constant torque, are synchronized to form a flagellar bundle. For each swimming mode, the flagellar motor applies a specific constant torque to produce that mode. No tumbles or transitions between swimming modes occur within a given simulation.
To mimic the experimental setting, each cell is positioned in a shear flow, such that the fluid velocity is zero in the focal plane where the cell is located. By imposing zero velocity in the focal plane, the flow field in our simulations corresponds to the co-moving frame of reference of the cell, i.e., the flow speed in the focal plane, as seen from the laboratory frame of reference, was subtracted. A uniform shear rate of $\gamma= 3.0$ s$^{-1}$ was applied in all simulations, unless stated otherwise (see Materials and Methods and Supplementary Material for further details).   

Figure~\ref{fig:tracks_model} shows trajectories of swimmers in different swimming modes and under shear flow conditions that we obtained from our numerical simulations.
The plane of focus, where the flow velocity is zero, is indicated in light blue.
The initial swimming direction was oriented within the plane of focus, aligned to the fluid flow in the adjacent planes.
Above the plane of focus, the fluid moves in line with the initial swimming direction (swimming with the flow), while below it moves against the initial swimming direction (swimming against the flow).
For each swimming mode, eight initial flagellar phases ($\phi$) were considered, and the resulting trajectories are shown in three views: perspective, side, and top.
 
Similar to the experimental results in Figure~\ref{fig:Angles_bulk}B, also in the simulations, a drift motion across the streamlines was observed for a pusher configuration, while the puller stably follows the streamlines over time intervals even longer than the average duration of run episodes observed in experiments (see Table~S3 in Supplementary Material). Interestingly, according to our simulations, swimmers in wrapped mode do not exhibit a directional bias but spread more rapidly in various directions as their trajectories are more strongly influenced by the initial flagellar phase than those of other swimming modes.

\vspace{2mm}
\noindent
\textbf{Pullers are stabilized under shear flow conditions.}
For {\pp} cells swimming in a fluid at rest the different swimming modes and the transitions between them have been studied in detail \cite{Hintsche2017e}.
When flagellar motors turn counterclockwise (CCW), cells propagate in pushing mode with the flagellar bundle propelling the cell body from behind.
Upon a switch to clockwise rotation, cells reverse their direction of motion and continue as pullers with the flagellar bundle pointing in the direction of motion.
In most cases, pull runs are short-lived and transition to the wrapped mode of locomotion, where cells propagate in a screw thread fashion with their helical bundle wrapped around the cell body \cite{Hintsche2017e, Alirezaeizanjani2020c}.
Numerical simulations suggest that an increase in the torque of the flagellar motors triggers the transition from pull to wrapped mode, where swimming speed and persistence of motion are decreased but motors continue to rotate in CW direction~\cite{10.1063/5.0228395}.
Once motors switch back to CCW rotation, swimming in wrapped mode ends and the cell changes back to pushing mode.

\begin{table}[t]
\begin{tabular}{>{\raggedright\arraybackslash}m{1.4 cm} 
                >{\centering\arraybackslash}m{1.3cm} 
                >{\centering\arraybackslash}m{1.3cm} 
                >{\centering\arraybackslash}m{1.3cm}  
                >{\centering\arraybackslash}m{1.7cm} } 
\toprule\multirow{2}{=}{shear rate $\gamma$, $s^{-1}$} & 
 \multicolumn{3}{c}{frequency of mode, \% } & 
 \multirow{2}{=} {ratio CW pull/wrap} \\
\cmidrule(l){2-4}
                          & push & pull & wrapped \\
\midrule
0.0 & 46.7     & 21.5  & 32.8 & 0.66\\
1.5 & 39.9     & 37.2  & 22.9 & 1.62\\
3.0 & 43.7     & 30.1  & 26.2 & 1.15\\
\bottomrule

\end{tabular}
\caption{\textbf{Ratio of swimming modes.} Contributions of the three swimming modes to the total number of runs in fluid at rest and in two different shear flows.
All data was recorded at a distance of $h = \SI{20}{\micro\metre}$ away from the surface.
\label{tab1}}
\end{table}

Here, we studied whether shear flows affect the transitions and thus the frequencies at which the individual swimming modes are observed.
Similar to earlier results~\cite{Pfeifer2022a}, a fraction of push runs (CCW rotation of the flagellar motors) between 40 and 50\% was observed in fluid at rest, as well as in two different shear flows, see Table~\ref{tab1}.
Within the fraction of pull and wrapped runs (CW rotation of the flagellar motors) we confirmed a dominance of the wrapped mode in resting fluid, see Table~\ref{tab1} and \cite{Pfeifer2022a}.
However, under shear flow conditions, the wrapped mode is reduced to less than 50\% within the population of CW runs and the pull mode dominates, see Table~\ref{tab1}.
The swimming speeds, by contrast, remained similar for cells in resting fluid and under shear condition, when measured at the same distance above the bottom surface of the channel, see Table~S4 in Supplementary Material.
These results suggest that filament wrapping becomes less likely in a fluid shear flow, thus stabilising swimmers in pulling mode, while push-pull transitions are less affected.

\vspace{2mm}
\noindent
\textbf{Critical torque for filament wrapping is increased in fluid shear flow.}

To confirm that filament wrapping is indeed affected under shear flow conditions, we performed numerical simulations
of cells transitioning from the pull to the wrapped mode under shear flow, with the flagellar motor rotating clockwise (CW), corresponding to negative values of the motor torque ($\tau$). Each cell in pull mode was initially positioned in the focal plane, where the fluid velocity was zero, i.e., numerical simulations were performed in the co-moving frame of reference, as described above. Various combinations of motor torque ($\tau$), shear rate ($\gamma$), and initial flagellar phase ($\phi$) were considered.
 
\begin{figure*}[t]
\centering    
\includegraphics[width=1.85\columnwidth]{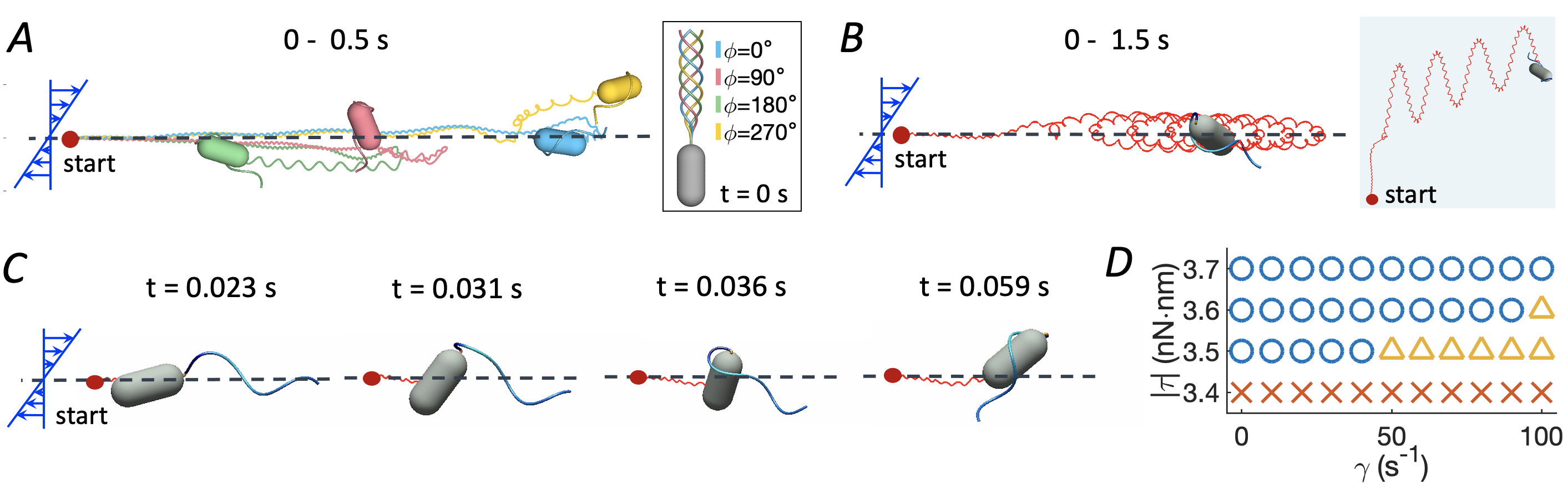}
\caption[Swimming modes, flow, bulk]{\textbf{Numerical simulations: wrapped mode formation under shear flow.} 
(A)~Tracks of swimmers undergoing a pull-to-wrap transition for four different initial flagellar phases $\phi$ (see inset). The dashed line indicates the focal plane, where the flow speed is zero. The applied torque and the shear rate are $\tau=-3.5$ nN$\cdot$nm and $\gamma =20$ s$^{-1}$, respectively.
(B)~Typical periodic trajectory of a wrapped swimmer at high shear rates ($\tau=-3.7$ nN$\cdot$nm, $\gamma=50$ s$^{-1}$, and $\phi=0^\circ$).
The left and right panels correspond to side and top views, respectively.
(C)~Snapshots showing time evolution of a cell's transition into the wrapped mode, 
$\tau=-4.0$ nN$\cdot$nm, ~$\gamma=50$ s$^{-1}$, and $\phi=90^\circ$.
(D)~Motor torque thresholds ($|\tau|$) for the formation of the wrapped mode at different shear rates ($\gamma$). Circles represent the wrapped mode, crosses represent the pull mode, and triangles represent cases where either mode may occur depending on the initial flagellar phase. See also the corresponding Videos~S4 to S6 in the Supplementary Information.
}
\label{fig:wrap_formation}
\end{figure*}

Figure~\ref{fig:wrap_formation}A displays four trajectories corresponding to four distinct initial flagellar phases ($\phi=0^\circ,~90^\circ,~180^\circ$, and $270^\circ$, see inset), while all other parameters were held constant.
It can be seen that the flagellar phase critically influences the turning angle during a transition from pull to wrapped mode.
As a result, the initial flagellar phase affects subsequent bacterial spreading in the wrapped mode. 
Depending on the interplay of turning angle and shear flow geometry, different complex trajectory patterns may emerge.
Figure~\ref{fig:wrap_formation}B illustrates, as an example, the long-time periodic trajectory of a wrapped swimmer observed at high shear rates. The left and right panels correspond to the side and top views, respectively, as in Figure~\ref{fig:tracks_model}. After the flagellum wrapped around the cell body, the wrapped-mode swimmer oscillated across the focal plane between the regions of opposing flow directions.
The detailed temporal evolution of a cell's transition into the wrapped mode can be seen in Figure~\ref{fig:wrap_formation}C for $\tau=-4.0$ nN$\cdot$nm, $\gamma=50$ s$^{-1}$, and $\phi=90^\circ$. At this torque magnitude, the flagellum buckles, causing the cell body to turn out of the focal plane, and hydrodynamic interactions lead to filament wrapping.

Finally, in Figure~\ref{fig:wrap_formation}D we show the resulting swimming mode as a function of motor torque and shear rate, starting from a pull mode swimmer with the motor rotating clockwise.
As has been shown in previous simulations in resting fluid, a critical motor torque is required to initiate filament wrapping~\cite{Park2022}.
This is also confirmed here, where the pull mode persists for small motor torques, independent of the shear rate.
Our simulations under flow conditions show that the critical torque that is required to initiate filament wrapping increases with increasing shear rate.
Note that, for each choice of torque and shear rate, eight initial flagellar phases uniformly distributed over $360^\circ$ were considered. The resulting swimming mode is generally independent of the initial phase, except near the threshold values, where different initial phases may lead to either pull or wrapped modes.
%
\section{Discussion}

The environmental conditions of bacterial habitats, such as complex geometries or fluid flows, may strongly affect bacterial locomotion and navigation.
In this work, we examined the impact of fluid shear flow on the motility pattern of the gram-negative soil bacterium {\pp}, a lophotrichously flagellated, rod-shaped swimmer that exhibits runs in push, pull, and wrapped mode and serves as a model for multi-mode bacterial swimming.

We showed that active swimming perturbed shear flow-induced alignment of the bacterial body axis with the flow direction.
This can be attributed to the turning maneuvers that occur between run phases and repeatedly randomize the direction of motion and, hence, the body orientation of the motile wild type cells.
Nevertheless, under shear flow conditions in the bulk fluid (20~$\mu$m above the bottom surface of the channel), swimming trajectories exhibit a clear tendency to align with the flow direction.
In contrast, close to a solid surface (5~$\mu$m above the bottom surface of the channel), flow alignment of the trajectories was no longer detectable.
We assume that this is due to the wall-induced reorientation of swimmers parallel to the solid surface, so that the rod-shaped cell bodies are less affected by the shear gradient perpendicular to the surface~\cite{Rubio2021}.

By considering runs in push, pull, and wrapped modes separately, we could show that alignment is more pronounced for pushers and pullers, while swimmers in wrapped mode showed no clear tendency, neither far nor near the interface.
This is presumably related to the compact shape of the wrapped mode as compared to the more elongated configurations of pushing and pulling swimmers (according to our fluorescence images, the aspect ratio of unwrapped swimmers is $\alpha \simeq 7.5$, while for wrapped swimmers it is only $\alpha \simeq 2.5$).
At the same time, a rheotactic drift across streamlines was observed for swimmers in push mode in agreement with earlier literature~\cite{Stocker_2011,Marcos2012b,Zhang2016b,Jing2020}.
Unexpectedly, however, only for the wrapped mode swimmer, an analogous effect was observed, while the trajectories of pullers were not affected over the time window of our observation.
Numerical simulations relying on realistic swimmer geometries confirmed the experimental observations for pushers and pullers, but did not show any directional preference for the wrapped mode swimmers.
The absence of a rheotactic drift of pullers in both our experimental and numerical data does not agree with our current understanding of rheotaxis as a consequence of chirality-induced cell re-orientation in a shear flow and remains an open question that requires further mechanistic studies.

As swimmers are floating along with the fluid flow, transitions between swimming modes and the associated directional changes may enhance the spreading efficiency.
For example, a run-and-reverse swimming strategy was shown to be advantageous for efficiently tracking small food sources in dynamic fluid environments~\cite{Guseva2025}.
Here, we demonstrated that the transitions between swimming modes are affected by fluid shear flow conditions.
Specifically, we observed that among the runs with CW rotating motors, the ratio between straight runs in pull mode and less persistent runs in wrapped mode is shifted in favor of pullers, indicating that the pull mode becomes more stable in a shear flow.
This is in line with earlier observations in {\it E.~coli} showing that run times and flagellar transitions are affected by the presence of a fluid flow~\cite{liu_upcoming_2021}.
Our numerical modeling results suggest that the pull mode is stabilized due to mechanical interactions between the flagella and the fluid shear flow, where different fluid velocities in the vicinity of the cell body make filament wrapping more difficult. Thus, an increased critical torque is required for wrapping in a shear flow, as demonstrated by our systematic numerical simulations. 
As a result, more pullers and less wrapped swimmers are observed in a shear flow as compared to liquid at rest.
Note also that this purely mechanical scenario does not involve any active regulation via receptor-mediated signaling.

Future studies of bacterial swimmers will focus on combining environments of complex geometry with hydrodynamic flows to mimic real-world habitats even more closely. 
They may also serve as a source of insight for the navigation of artificial microswimmers and may be combined with advanced techniques to induce time-dependent local flow fields by bio-compatible light-sensitive materials~\cite{Muraveva2024, Umlandt2025}.
Such experiments may also include mutant cell lines that allow for the adjustment of run lengths, stop events, and transitions from one swimming mode to another.
Here, constructing so-called non-tumbling mutants that propagate as push-only, pull-only, or wrapped-only swimmers will be particularly instructive to study the role of the different swimming modes within the complex environment.

\section{Conclusion}

In this study, we investigated how fluid shear flow influences the swimming behavior of \textit{Pseudomonas putida}, a soil bacterium known for its multi-mode swimming strategy.
Our experiments revealed that shear-induced alignment and rheotactic drift responses vary depending on the swimming mode.
We furthermore observed that flagellar wrapping becomes less effective under higher shear stress.
Numerical simulations based on realistic swimmer geometries support the experimental findings.
While still at an early stage, 
we hope to contribute with our findings to predicting the spreading of bacterial infections in complex surroundings as well as to the engineering of artificial microswimmers designed to navigate specific habitats.

\begin{acknowledgments}
V.M. acknowledges the kind hospitality of the Erwin Schrödinger Institute (ESI) in Vienna during the workshop "Non-equilibrium processes in Physics and Biology", where part of this project was productively discussed. We acknowledge fruitful discussions with Dr. Robert Großmann.
The research has been partially funded by the International Max Planck Research School on Multiscale Bio-Systems (V.M.) and by the Deutsche Forschungsgemeinschaft (DFG), project ID 443369470 -- BE~3978/13-1 (V.M., V.P., and C.B.) as well as project-ID 318763901 -- SFB1294 (A.D., V.P., and C.B.).
S.L. was supported by the NSF (CBET-2415406) and the Charles Phelps Taft Research Center at the University of Cincinnati, USA. Y.K. was supported by the National Research Foundation of Korea Grant funded by the Korean government (RS-2023-00247232). 
W.L. was supported in part by the National Institute for Mathematical Sciences Grant funded by the Korean government (B25910000) and in part by the National Research Foundation of Korea (NRF) grant funded by the Korea government (MSIT) (RS-2025-16071334).
During most of the period in which this work was carried out, J.P. was a member of the Institute for Advanced Study, supported by the AMIAS Fund, with access to IAS computing resources.

\end{acknowledgments}

\section*{Data availability}

The original code is publicly available at \cite{Datta_code2025}.
Any additional information required to reanalyze the data reported in this paper is available from the lead contact upon request.

\section*{Contributions}

VM designed and conducted experiments, analyzed experimental data and drafted the manuscript. AD designed the image analysis toolbox, analyzed experimental data and wrote the manuscript. SL and JP 
designed the numerical settings and wrote the manuscript. JP, YK and WL performed numerical simulations. JP processed and visualized numerical data. VP established the fluorescent staining procedure and generated the stator mutant. CB designed the research and wrote the manuscript. CB and SL  supervised the whole project. All authors reviewed and edited the manuscript, discussed the results and contributed to the final manuscript.

\section*{Declaration of interests}
The authors declare no competing interests.

\section*{SUPPLEMENTAL INFORMATION INDEX}

\begin{description}
  \item Figures S1-S2 and their legends in a PDF.
  \item Table S1-S4 and their legends in a PDF.

  \item Video S1-2. Examples of raw data and results of the image analysis toolbox processing as described in the main text. 
  \item Video S3. Numerical simulations: trajectories of cells swimming in a shear flow.
  \item Video S4. Numerical simulations: trajectories of cells transitioning from pull to wrapped under shear flow. 
  \item Video S5. Numerical simulation: long-time periodic trajectory of a wrapped swimmer at high shear rate.
  \item Video S6. Numerical simulation: time evolution of a cell’s transition into the wrapped mode under shear flow.
  
\end{description}


\bibliography{Manuscript_arXiv}

\end{document}


\title{SUPPLEMENTARY MATERIAL}
\author{
}
\date{\today}
\maketitle

\section{Semi-automatic analysis of trajectories}
\label{appsec:track_analysis}

\noindent

In this section, we discuss how the trajectories are obtained from microscope images and the criteria we use to identify the different swim modes. To track and distinguish cells in different swimming modes, bacteria are stained with fluorescent dyes (red for the body and green for the flagella), see Main Text and \cite{textbook}. Preliminary processing (signal splitting, frame cropping) of the fluorescent recordings is done with freely available \textit{ImageJ} software as described in the Main Text \cite{Hintsche2017}. For further tracking and analysis, the lab-made Python code \cite{Datta_code2025} is used, which is provided along with this manuscript. The workflow is also shown in Fig.~2 in the Main Text.

\subsection{Cell tracking}

\noindent
The raw images from the microscope are a stack of \textit{uint16} images consisting of two colored channels, namely the red and the green ones, which capture the staining of the cell body and the flagella bundle of the bacterial cells, respectively. As a first step, we split the image stacks into the red and green channels, thereby obtaining a pair of grey-scale image stacks. After we perform segmentation and tracking separately on the two image stacks to extract the trajectories corresponding to the center of masses of the segmented cell body and flagella of each moving cell. For the last step, we use the method described in publication~\cite{Theves2013}, which is based on Crocker's processing algorithm. Detailed description of the last one can be found in \cite{Crocker1996}.

\subsection{Distinction of swim-modes}

\begin{figure}
    \centering
    \includegraphics[width=0.5\linewidth]{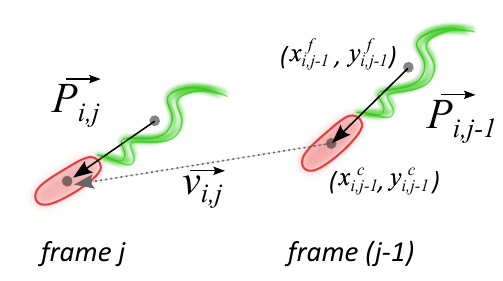}
    \caption{Schematic illustration of vectors of a single swimmer for the swim-mode detection.}
    \label{fig:vectore}
\end{figure}

\noindent
We match the coordinates of the cell trajectories (obtained from the red channel) and the flagella trajectories (obtained from the green channel). Based on a minimum displacement $\delta x$ and minimum time difference $\delta t$ between the coordinates of each point in the cell and flagella trajectories, we identify the pairs of trajectories which correspond to the individual bacteria. Then, for each identified pair and each frame of the image stacks, we define a vector from the coordinate of the center of mass of a cell  to that of the corresponding identified flagella bundle, which we call the pointing vector $\boldsymbol{P}_{i,j}$ and also the velocity vector $\boldsymbol{v}_{i,j}$ corresponding to the displacement of the cell from the previous frame (see Fig.~\ref{fig:vectore})., defined as follows:     

\begin{subequations}
\label{eq:runeq}
\begin{align}
\boldsymbol{P}_{i,j} & =  \left({x}_{i,j}^{f}-{x}_{i,j}^{c}\right) \hat{i} + \left({y}_{i,j}^{f}-{y}_{i,j}^{c}\right) \hat{j}, \label{eq:pointing} \\  
\boldsymbol{v}_{i,j} & =  \left({x}_{i,j}^{c}-{x}_{i-1,j}^{c}\right) \hat{i} + \left({y}_{i,j}^{c}-{y}_{i-1,j}^{c}\right) \hat{j}. \label{eq:velocity}
\end{align} 
\end{subequations}

Here, the index $i$ corresponds to the $i^{th}$ frame of the image, index $j$ corresponds to the $j^{th}$ pair of tracks (one bacterium), index $c$ corresponds to the coordinates of the center of mass of cell (obtained from red channel) and the index $f$ corresponds to the coordinates of the center of mass of flagella bundle (obtained from green channel). We determine the swim-mode corresponding to the $j^{th}$ bacterium in the 
$i^{th}$ frame by the inner product of $\boldsymbol{P}_{i,j}$ and $\boldsymbol{v}_{i,j}$ as shown in Table~\ref{table:swim-modes}. 

%
%

\begin{table}[h]
\centering
\begin{tabular}{ 
                >{\centering\arraybackslash}m{3.0cm} 
                >{\centering\arraybackslash}m{5.0cm} 
               } 
\toprule
        {swim-mode}&{criterion}\\
\midrule
push     & \large $\boldsymbol{P}_{i,j}.\boldsymbol{v}_{i,j} > 0$ \\[1ex]
pull     & \large $\boldsymbol{P}_{i,j}.\boldsymbol{v}_{i,j} < 0$  \\[1ex]
wrapped  & \large $\left|\boldsymbol{P}_{i,j}\right| < \epsilon$ \\[1ex]

\bottomrule
\end{tabular}
\caption{Criterion to identify swim-modes from a pair of trajectories in a particular frame. The threshold for cell body-flagella is distance $\epsilon=1.1$ $\mu$m for {\pp} cells.}
\label{table:swim-modes}
\end{table}

It is important to note here that the distinction is based on the parameters $\delta x$, $\delta t$ and $\epsilon$, which show little variability from image to image, especially for separate datasets due to the heterogeneity of the bacterial population and quality of recorded images. Further, to calculate the flow drift velocity, we analyze some non-swimming cells (pre-selected manually) in the cell suspension. Then, we perform segmentation and tracking on the red channel of the recorded images. Thereby, we obtain the drift velocity $\boldsymbol{v}_{i,j}^{\text{drift}}$ corresponding to the flow by taking the average of the velocities calculated from the derived trajectories. Then, we modify the detection criteria by taking the inner product of $\boldsymbol{P}_{i,j}$ and the relative velocity $\boldsymbol{v}_{i,j}^{\text{rel}}$, obtained by subtracting the flow velocity from the observed velocity. 

\begin{align}  
\boldsymbol{v}_{i,j}^{\text{rel}} & =  \boldsymbol{v}_{i,j} - \boldsymbol{v}_{i,j}^{\text{drift}}
\label{eq:relative_velocity}
\end{align} 

After, the same criteria is used to distinguish the swim-modes by replacing $\boldsymbol{v}_{i,j}$ by $\boldsymbol{v}_{i,j}^{\text{rel}}$ in Table~\ref{table:swim-modes}. 

Also, $\boldsymbol{v}_{i,j}^{\text{drift}}$ vector is used to define the orientations of the cells with respect to the flow streamlines. We defined the angle between the pointing vector $\boldsymbol{P}_{i,j}$ and fluid flow direction, calculated above, as $\psi$ angle (body's orientation), and the angle between the cell's velocity vector $\boldsymbol{v}_{i,j}$ and fluid flow direction as $\theta$ angle (the direction of the displacement), see also Fig.~1D in the Main Text. The $\psi$ angles (body's orientation) are defined only for unwrapped cells' configurations (push, pull and mutant non-motile cells).
\label{appsec:track_analysis}

\noindent


\section{Mathematical framework}

\begin{figure}[t!]
    \centering
\includegraphics[width=0.9\columnwidth]{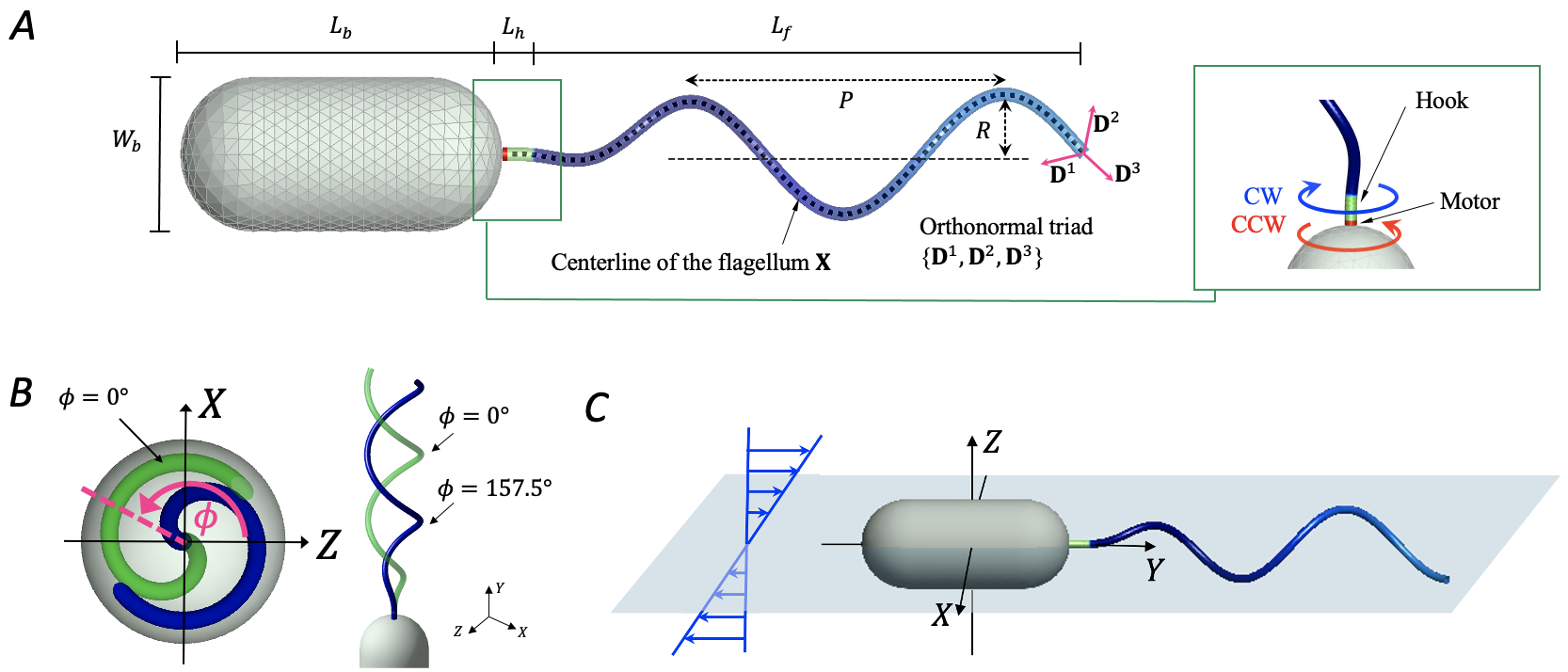}
    \caption{ 
    A) Schematic diagram of the numerical model organism. B)  Definition of the initial phase angle ($\phi$) of the flagellum. In the left panel, the initial phase angle $\phi$ is defined by counterclockwise rotation of the $\phi=0^\circ$ flagellum (green) about the helical axis. Flagella with initial phase angles of $0^\circ$ (green) and $157.5^\circ$ (navy) are shown in the right panel. 
    C) Initial setup under shear flow. The dark blue arrows indicate the direction and magnitude of the flow. The light blue panel represents the imaginary focal plane, where the fluid velocity is zero, corresponding to the experimental observation field. In the numerical simulation, flow velocities are computed by subtracting the experimental flow velocity from the measured velocity in the observation field.
    }
    \label{SI:numerical}
\end{figure}

{\it P. putida} consists of a rod-shaped cell body with a tuft of multiple flagella anchored at one pole, and it propels itself through fluid via flagellar rotation.  Our computational model is constructed based on the experimentally observed geometry of {\it P. putida} and incorporates the relevant biological data. For simplicity, the cell model includes a single polar flagellum representing a bundled configuration of multiple flagella. The initial setting and parameter values used in the simulations are provided in Fig.\ref{SI:numerical} and Table \ref{tab:parameters}, respectively.

Flagellar dynamics is modelled using Kirchhoff rod theory, while the cell body is treated as a rigid body. The forces and torques generated by flagellar rotation are transmitted to the surrounding fluid which is governed by the incompressible Stokes equations. To solve these coupled equations, we employ the immersed boundary framework that is useful to solve fluid-structure interaction (FSI) problems. This approach captures the hydrodynamic interactions between the cell and the fluid, enabling bidirectional coupling between the cell motion and the resulting fluid flow.
Importantly, the flagellar motor generates a constant torque, introduced as a control parameter, that drives flagellar rotation, while an equal and opposite torque is applied to the cell body, resulting in its counterrotation. The swimming motion is computed by enforcing the condition that the net force and torque exerted on the surrounding fluid are both zero. Additionally, in the presence of shear flow, another parameter is introduced to control the imposed shear rate. 
We begin by describing the motion of a model cell composed of a rod-shaped cell body and a single left-handed superflagellum, formulated in Lagrangian coordinates. Subsequently, we present the governing fluid equations using the regularized Stokeslet formulation expressed in Eulerian coordinates. A detailed description of the method can be found in~\cite{Lim2013,Lim2021} and therein; key features of the approach are summarized below.

The cell body is modeled as a spherocylinder and discretized into $n_{\rm b}$ points on its surface. It is represented in two forms, ${\bf X}_{i}^{\rm b}(t)$ and ${\bf Y}_{i}^{\rm b}(t)$, for $i=1,2,...,n_{\rm b}.$ The former corresponds to the physical body that interacts with the surrounding fluid, while the latter represents a rigid body that does not directly interact with the fluid. Each pair of corresponding points ${\bf X}_{i}^{\rm b}(t)$ and ${\bf Y}_{i}^{\rm b}(t)$ is connected via a stiff spring which generates a restoring force of the form
\begin{equation}
{\bf F}_{i}^{b} (t) = K({\bf X}_{i}^{\rm b}(t) - {\bf Y}_{i}^{\rm b}(t)),
\end{equation}
where $K$ is a large stiffness constant that ensures the two representations remain close to each other. The force ${\bf F}_{i}^{b} (t)$ is applied to the rigid body at ${\bf Y}_{i}^{\rm b}(t)$, while an equal and opposite force $-{\bf F}_{i}^{b} (t)$ is exerted on the fluid at ${\bf X}_{i}^{\rm b}(t)$. 
The reference configuration of the rigid cell body is defined by a set of time-independent points ${\bf Z}_{i}$ that satisfy $\sum_{i=1}^{n_{\rm b}} {\bf Z}_{i}={\bf 0}$. 
The position of the rigid body at time $t$ is then given by
\begin{equation}
{\bf Y}_{i}^{\rm b}(t) = {\bf C} (t) + \mathcal{R}(t){\bf Z}_{i},
\end{equation}
where ${\bf C}(t)$ is the centroid of the rigid cell body and $\mathcal{R}(t)$ is a rotation matrix.

Let $\mb f^{\rm b}(t)$ and $\mb n^{\rm b}(t)$ denote the external force and torque, respectively, acting on the cell body, excluding contributions from the spring connections. Then, the balance of linear and angular momentum for the rigid body is given by 
\begin{align}
    0 = \mb f^{\rm b}(t)  + \sum^{n_{\rm b}}_{i=1}{\mb F}^{\rm b}_i(t),\hspace{1cm}
    0 =\mb n^{\rm b}(t)  +   \sum^{n_{\rm b}}_{i=1}(\mathcal R(t) {\mb Z}_i)\times {\mb F}^{\rm b}_i(t).\label{eq:rigid_motion}
\end{align}
Given $\mb f^{\rm b}(t)$, $\mb n^{\rm b}(t)$ and ${\bf X}_{i}^{\rm b}(t)$ at each time $t$, we can compute ${\bf C}(t)$, $\mathcal{R}(t)$, and thus ${\bf Y}_{i}^{\rm b}(t)$ by solving the above system approximately~\cite{Lim2021}. The accuracy of the approximation improves with increasing spring stiffness $K$.

The motion of a flagellum is described by Kirchhoff rod theory. A flagellum is represented as a tapered helical curve $\mb X(s,t)$ together with orthonormal triads $\{\mb D^1(s,t),\mb D^2(s,t),\mb D^3(s,t)\}$ along the curve. This curve represents the centerline of the flagellum and the triads show the degree of bending and twisting. 
Let ${\mb F(s,t)}$ and ${\mb N(s,t)}$ be the force and the moment, respectively, which are transmitted across a section of the rod at the Lagrangian coordinate $s$ at time $t$. Let ${\bf f}(s,t)$ and ${\bf n}(s,t)$ be the applied force and the torque densities, respectively. Then the linear and angular momentum balance equations are described as follows:
\begin{equation}\label{flag balance equation}
    0 = {\bf f} + \dfrac{\partial {\bf F}}{\partial s}, \qquad 0 = {\bf n}+ \dfrac{\partial {\bf N}}{\partial s} + \dfrac{\partial {\bf X}}{\partial s} \times {\bf F}.
\end{equation}
Here, the internal force ${\bf F}$ and torque ${\bf N}$ can be written as follows:
\begin{equation}\label{relation}
    {\bf F} = \sum_{i=1}^{3} b_{i} \Big( {\bf D}^{i}\cdot \dfrac{\partial {\bf X}}{\partial s} - \delta_{3i} \Big) {\bf D}^{i}, \qquad {\bf N}= \sum_{i=1}^{3} a_i \Big( \dfrac{\partial {\bf D}^{j}}{\partial s} \cdot {\bf D}^{k} - \kappa_{i} \Big) {\bf D}^{i},
\end{equation}
where $\delta_{3i}$ is the Kronecker delta, $(i,j,k)$ is a cyclic permutation of $(1,2,3)$, and $\kappa_{i}$ describes the intrinsic curvature ($\kappa_1$ and $\kappa_2$) and twist ($\kappa_3)$. The constitutive relations can be derived from a variational derivative of the elastic energy functional for the unconstrained version of the Kirchhoff rod,
\begin{equation}
    E = \dfrac{1}{2} \Big[ 
    \sum_{i=1}^{3} a_{i} \Big( \dfrac{\partial {\bf D}^{j}}{ \partial s} \cdot {\bf D}^{k} - \kappa_{i} \Big)^{2}
     + \sum_{i=1}^{3} b_{i} \Big( {\bf D}^{i} \cdot \dfrac{\partial {\bf X}}{\partial s}- \delta_{3i}\Big)^{2}
    \Big] ds,
\end{equation}
where the parameters $a_{1}$ and $a_{2}$ are bending moduli, $a_3$ is the twist modulus, $b_1$ and $b_2$ are shearing moduli, and $b_3$ is the stretching modulus.
To model the compliant hook, the bending modulus of the hook $a^{\rm hook}$ is reduced by a factor of 25 compared to that of the flagellum, reflecting its increased flexibility.

The tapered end of the flagellum is embedded into the cell body at the pole, with the flagellar axis aligned with the normal direction $\mathbf{E}(t)$ to the cell surface. The flagellar motor generates a constant torque $\tau$ in this normal direction at the motor point $\mathbf{X}(s_{\rm mot}, t),$ given by
${\mathbf{N}}(s_{\rm mot},t)=-\tau\mathbf{E}(t)$. 
Positive values of $\tau$ correspond to counterclockwise rotation, while negative values correspond to clockwise rotation. This motor torque drives the rotation of the flagellum, enabling the cell to swim through the surrounding fluid.

Finally, the cell motion described above is coupled with the surrounding fluid which is governed by the incompressible Stokes equations:
 \begin{align}\label{eq:Stokes}
    0 = -\nabla p + \mu \Delta \mb{u} + \mb g,  \qquad 0 = \nabla \cdot \mb{u},
\end{align}
where the fluid velocity $\mathbf{u}$ and pressure $p$ are unknown functions of the Cartesian coordinates $\mathbf{x}$ and time $t$, and $\mu$ is the dynamic viscosity of the fluid.
The regularized force density $\mathbf{g}$ applied to the fluid by the immersed cell is given by:
\begin{align}\label{eq:g}
\begin{aligned}
    \mb{g}(\mb x, t) &= \int_{0}^{L} \big[- \mb f(s,t)\big] \varphi_{\varepsilon}\big(\mb x - \mb X(s,t)  \big) ds +  \frac{1}{2} \nabla \times \int_{0}^{L} \big[ - \mb n(s,t) 
    \big] \varphi_{\varepsilon} \big( \mb x - \mb X(s,t) \big)  ds \\
    & \quad +\sum_{i=1}^{n_{\rm b}}
    \int_{L'}^{L} \big[- \mb f_i^{\rm r} (s,t)  \big] \varphi_{\varepsilon}\big(\mb x - \mb X(s,t)  \big)ds + \sum_{i=1}^{n_{\rm b}}
    \big[ \int_{L'}^{L}\mb f_i^{
    \rm r} (s,t) ds \big] \varphi_{\varepsilon}\big(\mb x - \mb X_{i}^{\rm b}(t)\big)    \\
    &\quad + \sum_{i=1}^{n_{\rm b}}(-{\mb F}^{\rm b}_i(t)) \varphi_{\varepsilon}(\mb x - \mb X_i^{\rm b}(t)).
\end{aligned}
\end{align}
The first and second terms on the right-hand side represent the force and torque, respectively, exerted by the flagellum, and the third and fourth terms correspond to a short-range repulsive force that prevents contact between the flagellum and the cell body. Note that the repulsive force is not activated near the motor end of the filament by setting the lower limit  $L'$ of the integral to a value greater than zero. The fifth term accounts for the force exerted by the cell body, as discussed above. 
The bell-shaped blob function $\varphi_{\varepsilon}$, with infinite support, is defined as  
\begin{equation}
\varphi_\varepsilon(\mathbf{r}) = \frac{15\varepsilon^4}{8\pi (\lVert \mathbf{r} \rVert^2 + \varepsilon^2)^{7/2}},
\end{equation}
where $\varepsilon$ is the regularization parameter. The blob function satisfies $\int_{\mathbb{R}^3} \varphi_{\varepsilon}({\bf r}) d{\bf r}=1$, enabling smooth transfer of forces and torques from Lagrangian points of the model cell to the Eulerian fluid domain. 
The motion of the cell, including the body points ${\bf X}_{i}^{\rm b}$ and the flagellar configuration \{${\bf X}, {\bf D}^{1}, {\bf D}^{2}, {\bf D}^{3}\}$, is governed by the following equations:
\begin{equation}
\begin{gathered}
    \frac{\partial \mb X}{\partial t}(s,t) = \mb u \big(\mb X(s,t),t \big) -  \frac{1}{\alpha_1} \Big[ \mb f (s,t) + \sum_{i=1}^{n_{\rm b}} \mb f_{i}^{\rm r} (s,t) \Big],\\  
  \frac{\partial \mb X_i^{\rm b}}{\partial t}(t) = \mb u(\mb X_i^{\rm b} (t),t) - \frac{1}{\alpha_2}  \Big[ \mb F_i^{\rm b}(t) - \int_{L'}^{L} \mb f_{i}^{\rm r}(s,t) ds \Big], \qquad i = 1, ..., n_{\rm b},\\
    \frac{\partial \mb D^{j}}{\partial t}(s,t) = \Big[ \mb w \big(\mb X(s,t),t \big)  - \frac{1}{\beta} n_{3}(s,t) \mb D^{3}(s,t)  \Big] \times \mb D^{j}(s,t) , \qquad j=1,2,3,
    \label{update}
    \end{gathered}
    \end{equation}
where the angular velocity of the fluid is given as ${\bf w} = \dfrac{1}{2} \nabla \times {\bf u}({\bf x},t)$. The parameters $\alpha_i$'s and $\beta$ are translational and rotational drag coefficients, respectively, and $n_{3}(s,t)$ is the axial component of torque density $\mb n$.

To create shear flow,  we impose a linear shear profile in the $y$-direction, with the shear velocity field defined as
\begin{align}
   \mathbf{u}_{{\rm shear}}(\mathbf{x}) = (0, \gamma z, 0),
\end{align}
where $\mathbf{x}=(x,y,z)$ and $\gamma$ denotes the shear rate. The corresponding angular velocity field associated with this shear flow is given by 
\begin{equation}
\mathbf{w}_{\rm shear}(\mathbf{x}) = \frac{1}{2} \nabla  \times {\bf u}_{\rm shear} = (-\frac{\gamma}{2}, 0 ,0). 
\end{equation}
Both the translational velocity $\mathbf{u}_{{\rm shear}}(\mathbf{x})$ and the angular velocity $\mathbf{w}_{\rm shear}(\mathbf{x})$ are added directly to the velocities of the immersed cell as specified in Eq. \ref{update}. 


\begin{table}[thb]
\centering\renewcommand\cellalign{lc}
\setcellgapes{3pt}\makegapedcells
\caption{Computational and physical parameters} 
\label{tab:parameters}
\resizebox{1.0\textwidth}{!}{%
\begin{tabular}{ll}
\toprule[0.5mm]
\multicolumn{1}{c}{\textbf{Parameter (symbol)}} & \multicolumn{1}{c}{\textbf{Value}} \\
\midrule[0.3mm] 
 Mesh width for flagellum ($\Delta s$) & \SI{0.0303}{\micro\metre}  \\
  Helical radius of filament ($R$) & \SI{0.35}{\micro\metre} \\
  Helical pitch of filament ($P$) & \SI{2.0}{\micro\metre} \\
  Axial length of helical filament  ($L_{\rm f}$) & \SI{3.5}{\micro\metre} \\
  Curvilinear length of filament ($\ell$) &\SI{4.99}{\micro\metre}\\
  Intrinsic curvature of filament ($\kappa_1=\kappa_2$) & \SI{1.5637}{\per\micro\metre} \\
  Intrinsic twist of filament ($\kappa_3$) & 
  \SI{1.4222}{\per\micro\metre} \\
  Bending modulus of filament ($a_1=a_2=a$) & \SI{0.003}{\gram \micro\metre^3 \second^{-2}} \\
  Twist modulus of filament ($a_{3}$) & \SI{0.003}{\gram \micro\metre^3 \second^{-2}} \\
  Shear modulus ($b_1 = b_2$) & \SI{2}{\gram \micro\metre \second^{-2}} \\
  Stretch modulus ($b_3$) & \SI{2}{\gram \micro\metre \second^{-2}} \\
  Length of hook ($L_{\rm h}$) & 4$\Delta s$   \\
  Bending modulus of hook ($a_{1,2}^{\rm hook}=a^{\rm hook}=a/25$) &  
  \SI{0.0012}{\gram \micro\metre^3 \second^{-2}} \\
  Twist modulus of hook ($a_{3}^{\rm hook}=a_{3}$)  & \SI{0.003}{\gram \micro\metre^3 \second^{-2}} \\
  Torque ($\tau$) & 2.0 (push), -2.0 (pull), -4.0 (wrapped) nN$\cdot$nm \\
  [1mm] \hline \\[-3mm]
 Cell body length ($L_{\text{b}}$) & \SI{2.2}{\micro\metre} \\
  Cell body width ($W_{\text{b}}$) & \SI{0.9}{\micro\metre} \\ 
  Penalty parameter for restoring force ($K$) & \SI{2.0}{\gram \second^{-2}} \\
  Number of material points on the cell body ($n_{\text{b}}$)  & 770
 \\[1mm] \hline \\[-3mm]
   Fluid viscosity ($\mu$) &  \SI{1e-4}{\gram (\micro\metre\second)^{-1}} \\
  Regularization parameter ($\varepsilon$) & $3\Delta s$  \\
  Translational drag coefficient ($\alpha_1$) &
  \SI{1.3221e-4}{\gram \second^{-1}} \\
Translational drag coefficient ($\alpha_2$) & 
\SI{4.0e-6}{\gram \micro \metre \second^{-1}}  \\
 Rotational drag coefficient ($\beta$) &  \SI{1.3221e-4}{\gram \second^{-1}} \\
 Shear rate ($\gamma$) & $0 \sim 100$ s$^{-1}$
 \\
  Time step ($\Delta t$) & $2\times 10^{-8}$ s
 \\
\bottomrule[0.5mm]
\end{tabular}%
}
\end{table}

\FloatBarrier

\section{Section: Supplemental experimental results}
In this part, we discuss how the orientation of bacterial cells is defined with respect to fluid flow streamlines. Also, we consider the role of motility in the orientation and show the re-orientation process of the individual swimming cells.

\subsection{Swimmers' velocities and run time}
To demonstrate that swimming bacteria do not change their motility in response to local shear, the swimming speeds are measured under flow conditions and in the rest fluid. Velocities are comparable for cell populations at the same distance above the channel surface.

\begin{table} [h]
\centering
\begin{tabular}{>{\raggedright\arraybackslash}m{1.8 cm} 
                >{\centering\arraybackslash}m{1.7cm} 
                >{\centering\arraybackslash}m{1.7cm} 
                >{\centering\arraybackslash}m{1.7cm}} 
\toprule
\multirow{2}{=}{conditions} & 
 \multicolumn{3}{c}{run time, $s$ } \\
 
\cmidrule(l){2-4}
                          & push & pull & wrapped \\
\multicolumn{4}{l}{\cellcolor{gray!20}\textit{bulk:}} \\ 
no flow & 0.30     & 0.39  & 0.56 \\
flow    & 0.34     & 0.43  & 0.54 \\
\multicolumn{4}{l}{\cellcolor{gray!20}\textit{surface:}} \\ 
no flow & 0.39     & 0.25  & 0.34 \\
flow    & 0.31     & 0.62  & 0.49 \\
\bottomrule
\end{tabular}

\caption{\textbf{The mean run time in different swimming modes}. By italic font, the measurement focal area is specified, where \textit{bulk} is related to the $h = \SI{20}{\micro\metre}$ and \textit{surface} to the $h=\SI{5}{\micro\metre}$ above the bottom surface of the channel.}
\label{run_times}
\end{table}


\begin{table} [h]
\centering
\begin{tabular}{>{\raggedright\arraybackslash}m{1.6 cm} 
                >{\centering\arraybackslash}m{1.5cm} 
                >{\centering\arraybackslash}m{1.5cm} 
                >{\centering\arraybackslash}m{1.5cm}} 
\toprule
\multirow{2}{=}{conditions} & 
 \multicolumn{3}{c}{swimming velocity, $ \mu m/s$ } \\
\cmidrule(l){2-4}
                          & push & pull & wrapped \\
\midrule
no flow  & 25.8     & 16.6  & 17.2 \\
flow     & 23.2     & 17.0  & 13.4 \\

\bottomrule
\end{tabular}
\caption{\textbf{The mean population velocities in different swimming modes, $h = \SI{20}{\micro\metre}$}. }
\label{velocities}
\end{table}

\FloatBarrier

\subsection{Cell body orientations and velocity vectors}
We define two types of bacterial cell orientation angles: $\psi$ corresponds to the angle between the pointing vector (vector pointed from the center of mass of the flagella to that of the cell body) and the direction of flow (drift velocity). On the other hand, $\theta$ corresponds to the angle between the direction of the displacement (by connecting the cell body positions from the previous frame to the current frame) and the direction of flow, see also Fig.~1D, in the Main Text. The $\psi$ angles are defined for push and pull modes.

\begin{figure}[t] 
\centering    
\includegraphics[width=0.8\columnwidth]{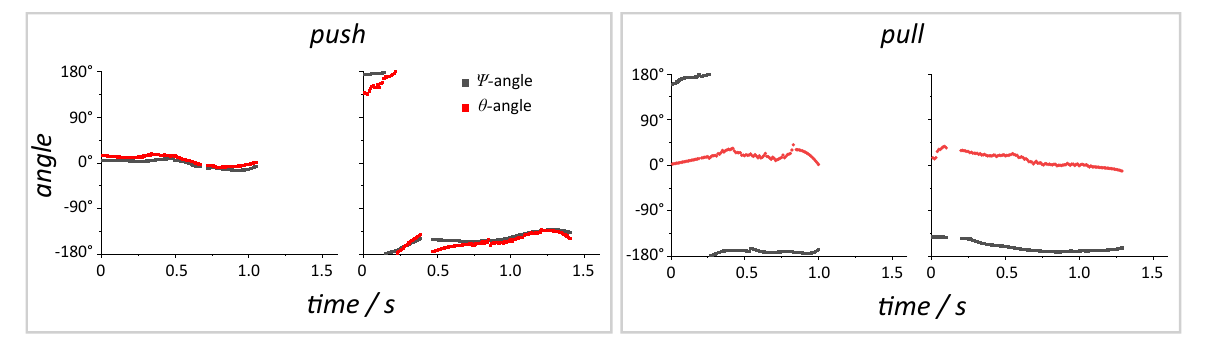}
\caption[psi and tetha]{\textbf{Typical examples of the cells orientation with time.} Cells' bodies orientations($\psi$ angles) and direction of forward swimming ($\theta$ angles) with respect on flow, shear rate 1.5 s-1, $h= \SI {20}{\micro\metre}$. The gaps in the tracking cause to the loss the one of the signals (red or green) of the swimmer.}
\label{SI_fig:psi_and_tetha}
\end{figure}

The angles as a function of time for several typical swimmers are shown in Fig.\ref{SI_fig:psi_and_tetha}. As expected, for the push configuration, $\psi$ and $\theta$ angles values are almost the same, the cell body is oriented similarly to the direction of the swimming motion. The difference between the direction of motion and the cell body orientation is around 180$^{\circ}$ for the pulled configuration.

Throughout the cell population, the cell displacement direction ($\psi$ angle) can be assigned with cell body orientation ($\theta$  angle) with respect to flow. However, for individual tracks, the time evolution of the angles might be slightly different, see Fig.\ref{SI_fig:psi_and_tetha}. For the wrapped swimming mode, the $\psi$ angle is not defined, and we discuss only the $\theta$ angle in the Main Text.

%
%


%
%

%
%
\subsection{Dynamics of cell reorientation} 

\begin{figure*}[h] 
\centering    
\includegraphics[width=0.8\columnwidth]{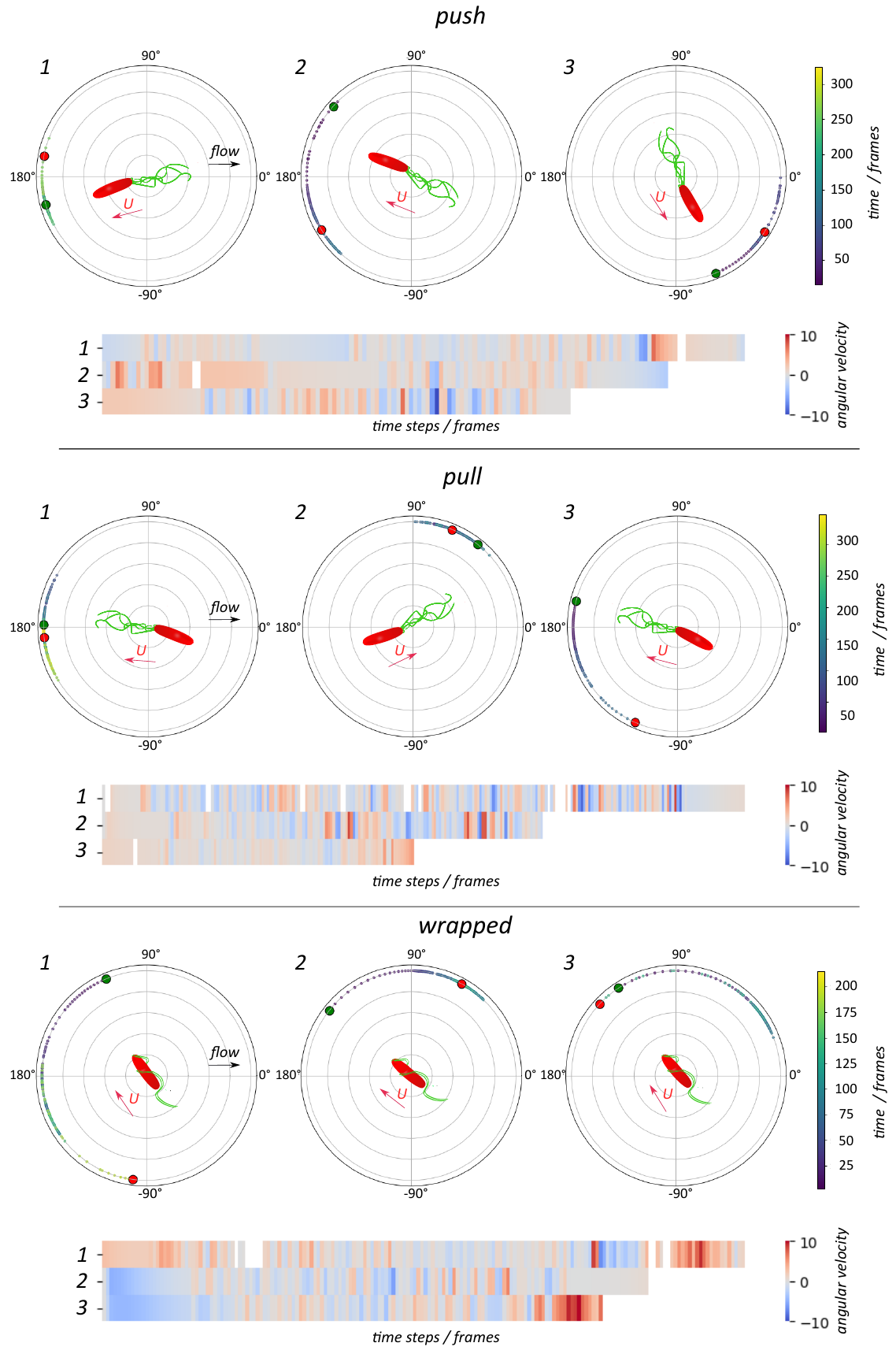}
\caption[Angles Ind]{\textbf{Orientations of velocity vectors($\psi$ angles) and angular velocities of individual cells in different swimming modes.} Results are shown for the typical long episodes in bulk. Colorbar representing the time, time resolution 100 fps, the green dots indicate the angle at the beginning of the episode, the red dots point to the end of it. Heatmap is shown for the filtered angular velocities between +10$^{\circ}$ and -10$^{\circ}$. Shear rate 1.5 s-1, $h = \SI{20}{\micro\metre}$.}
\label{SI_fig:ind_angles}
\end{figure*}

To show the details of the cells' reorientation process, the $\theta$ angles with time are considered for the long swimming episodes, see Fig.\ref{SI_fig:ind_angles}. The distance between single dots corresponds to the velocities of the angular changes. For the wrapped configuration, the angular velocities are higher in comparison with the unwrapped modes (see Fig.~\ref{SI_fig:ind_angles}, Heat-maps) and the amplitude of the angular changes along the tracks are bigger (see green and red dots in plots). For the unwrapped modes, the reorientations around 180$^{\circ}$ and -180$^{\circ}$ are slower across streamline orientations.
The re-orientation of the swimming locomotion direction is not persistent. The competition between the alignment and chirality-driven drift-side drift can cause it at the same time as thermal noise.

%

